\documentstyle[twocolumn,prl,aps,psfig]{revtex}

\catcode`\@=11

\def\maketitle2{\par 
\begingroup
\let\cite\@bylinecite
\def\thefootnote{\fnsymbol{footnote}}%
\twocolumn[\@maketitle2\vskip2pc]%
\thispagestyle{plain}\@thanks
\endgroup
\def\thefootnote{\arabic{footnote}}%
\setcounter{footnote}{0}%
\let\maketitle2\relax \let\@maketitle2\relax
\let\@thanks\relax \let\@authoraddress\relax \let\@title\relax
\let\@date\relax \let\thanks\relax \let\@abstract\relax 
\let\@pacs\relax}

\def\abstract#1{\gdef\@abstract{{\par 
\bgroup
\ifdim\prevdepth=-1000pt \prevdepth0pt\fi
\hsize\columnwidth
\dimen0=-\prevdepth \advance\dimen0 by17.5pt \nointerlineskip
\small\vrule width 0pt height\dimen0 \relax}{~~}#1\egroup}}

\def\pacs#1{\gdef\@pacs{{\par 
\bgroup
\hsize\columnwidth \parindent0pt
\ifdim\prevdepth=-1000pt \prevdepth0pt\fi
\dimen0=-\prevdepth \advance\dimen0 by20pt\nointerlineskip
\egroup} PACS numbers:~#1}}

\def\@maketitle2{
\@preprint
\@title
\ifdim\prevdepth=-1000pt \prevdepth0pt\fi
\@authoraddress
\@date
\begin{list}{}{\leftmargin=0.10753\textwidth \rightmargin=\leftmargin
\itemsep=1pc\partopsep=-1pc}
\item\@abstract
\item\@pacs
\end{list}
}

\catcode`\@=12

\begin{document}
\draft
\title{Anomalous transport and quantum-classical correspondence}
\author{Bala Sundaram\mbox{}$^{1}$ and G. M. Zaslavsky\mbox{}$^{2}$}
\address{\mbox{}$^1$ Department of Mathematics, CSI-CUNY, Staten Island,
NY 10314}
\address{\mbox{}$^2$ Courant Institute of Mathematical Sciences, New
  York University, New York, New York 10012 and \\
Department of Physics, New York University, 2-4 Washington Place, New
York, New York 10003}
\date{28 February 1998}
\abstract
{\small{We present evidence that anomalous transport in the classical
standard map results in strong enhancement of fluctuations in
the localization length of quasienergy states in the corresponding
quantum dynamics. This generic effect occurs
even far from the semiclassical limit and reflects the interplay of
local and global quantum suppression mechanisms of classically chaotic
dynamics. Possible experimental scenarios are also discussed.}}

\maketitle2

\narrowtext

It is generally accepted that quantum mechanics suppresses chaotic classical
motion. Numerous studies have identified
mechanisms for suppression which, for our purposes, fall into two
broad classes.  
One class is exemplified by `dynamical localization' where 
quantum eigenstates are localized in an action variable like momentum 
despite the deterministic diffusion exhibited in the limiting classical 
dynamics \cite{Casati95}. This effect is analogous 
to the insulator phase or Anderson localization in tight-binding 
models \cite{Fishman82} and, in one-dimension, the localized wavefunction 
has a characteristic exponential form. An associated 
scale is the localization length $\xi$ {\it which is related to the classical 
diffusion constant}, an important fact for the work reported here. There
also exists a demarcating timescale $t^*$ beyond which quantum 
and classical dynamics deviate. Despite a few counterexamples, dynamical 
localization provides a `global' mechanism for quantum suppression.

The second class can be motivated by the fast deviation of wavepacket 
dynamics from the classical motion even in the semiclassical limit. This
effect is contained in the logarithmic break time 
$\tau_{\hbar}=\ln{(I/\hbar)}/\Gamma$ for the breakdown of quantum-classical
correspondence~\cite{BerZas}, relative to a characteristic Lyapunov
exponent $\Gamma$ and action $I$. However, the quantum dynamics retains
features of the classical dynamics for times beyond this estimate. The
phenomenon of `scarring' \cite{Heller84,HellPT} refers to quantum
coherences associated with local, unstable and marginally stable, classical invariant
structures. Though there are still many open questions, this effect
has been numerically and experimentally observed in a wide
variety of strongly coupled quantum systems. 

In this letter, we present evidence for
another example of `classical persistence' related to anomalous
diffusion and accelerator modes in chaotic dynamics. Recently, 
there has been considerable activity in the area of anomalous
classical transport~\cite{george,BenK97,ishiz91}. What is of particular
relevance to the present work is that even in an established  paradigm 
like the standard map, described by $q_{n+1} = q_{n} + p_{n+1}$ and
$p_{n+1} = p_{n} + K \sin{q_n}\;,$ some surprising new results were reported.

\begin{figure}
\begin{center}
\leavevmode
\psfig{figure=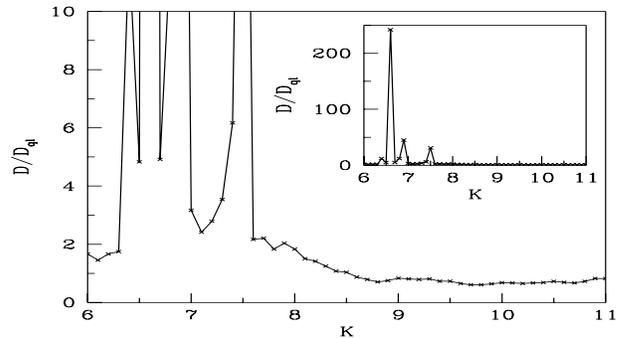,width=3.0in,height=1.8in}
\end{center}
\caption{
Diffusion coefficient $D$ for the standard map, normalized to
$D_{ql}=K^2/2$, as a function of
$K$. The deviations from
the quasilinear prediction span a wide window of $K$ and, as
seen from the inset, can be large at select values of $K$, even far
from the bifurcation point $K=2\pi$.}
\label{fig1}
\end{figure}

In the limit of large $K$, it is well-known that the classical
standard map dynamics is diffusive in the action variable $p$ with 
diffusion constant $<p^2>/t=K^2/2 \equiv D_{ql}$. What is 
also established is that 
away from this limit, the diffusion constant 
exhibits peaks with changing $K$ and varies in time~\cite{ishiz91}.
Recent results \cite{george,BenK97} with improved precision and 
detail show that there are values of $K$ where the 
effective diffusion constant (over some 
time) can be different from $D_{ql}$ by many orders of
magnitude. Figure~\ref{fig1} shows these windows of anomalous
diffusion which are not very narrow and exhibit sub-structure 
in each of the peaks.

It was shown that there is no diffusion at all and that the real
random walk process corresponds to a L\'{e}vy-like wandering 
characterized by a transport exponent $\mu$ 
given by $<p^2> \approx t^{\mu}$,
where $\mu >1$. $\mu$ varies with $K$ and has the
`normal' value $\mu=1$ only at special values of $K$. Superdiffusion
occurs for $\mu >1$ leading to anomalous growth in momentum.
It was explicitly shown~\cite{george} that this
could result from an island hierarchy with a peculiar topological structure 
near the `accelerator' modes that appear at special values 
of $K$. Near these values, classical trajectories stick to the
boundaries of these islands which, in turn, leads to flights of
arbitrary length. The net result is strong intermittency and
superdiffusion. 

\begin{figure}
\begin{center}
\leavevmode
\psfig{figure=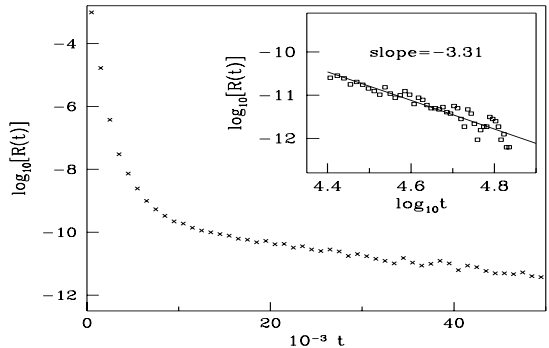,width=3.0in,height=1.8in}
\end{center}
\caption{
Distribution of Poincar\'{e} cycles. The inset clearly 
displays the power-law tail of the distribution.}
\label{fig2}
\end{figure}

Another signature of anomalous transport results from the fact that
a classical trajectory originating in a region of phase space recurs
infinitely often. From the set of recurrence times
$\{\tau_j \},\; j=0,1,\cdots $, 
Poincar\'{e} cycles $\{ t_j \}=\{ \tau_{j+1}-\tau_j \}$
and an associated probability distribution function $R(t)$ can be
constructed. In the case of `perfect' mixing and normal diffusion, 
$R(t)$ is strictly Poissonian
whereas it exhibits power-like asymptotics for anomalous kinetics.
This classical characteristic of anomalous transport is
illustrated in Fig.~\ref{fig2}, and
is important in the corresponding quantum dynamics as well.

We raise several questions, arising from the
classical anomalous 
diffusion, in the corresponding quantum dynamics of the standard
map or, equivalently, the `delta-kicked
rotor',
$H=p^2/2 + K\cos{q}\sum_n \delta(t-n) \;.$
The scale length $\xi$ for dynamical localization is related to the classical 
diffusion constant and even follows the variations in $D$ with
$K$~\cite{dima}. Thus, we anticipate that the strong enhancement in $D$
resulting from anomalous diffusion should also be reflected in $\xi$. The
nature of dynamical localization in these special windows is a related
question. Further, as small local structures
in the classical phase space lead to the superdiffusion, these
issues directly pertain to the interplay of local and global 
aspects of quantum suppression. This idea is reinforced by the
enhanced diffusion that ought to be visible in the 
quantum dynamics even far from the
semiclassical limit, where scarring is significant. 
Other authors have
considered quantum dynamics in the presence of anomalous 
transport\cite{NTexas,Geisel} though in a different regime. 
We concentrate exclusively on the situation
where the quantization scale $2\pi\hbar$ is much larger than the size
of the islands. Thus, any coherences associated with the islands would
be due to `scarred' quantum states. 

Consider the evolution of 
$<p^2>$ with time $t$ (measured as the number of kicks) in the quantum
dynamics for two values of the classical stochasticity parameter $K$.
The computation uses fast-Fourier
transforms to evolve a plane wave initial state under 
repeated application of the single-kick
evolution operator 
\begin{equation}
U = \exp{(-ip^2/2\hbar)} \exp{(-iKcos(q)/\hbar)}\;.
\end{equation}
The expectation values are then computed from the time-evolved wavefunction.
The value $K_c=6.908745\cdots$ was found~\cite{george} to be a critical
value in a wide parametric window dominated by anomalous transport. We
illustrate this regime by considering $K^*=6.905$ and contrast it
with $K=11$, where the effective diffusion coefficient is actually  below
the quasilinear value $D_{ql}$ (seen in Fig.~\ref{fig1}). The
quantization scale $\hbar=2\pi a$,
where several irrational values of $a$ in the range $0.04-1$ were considered.
In the quasilinear limit, the quantum diffusion is expected to saturate
at $t^* = \alpha D_{ql}/\hbar^2$ where $\alpha$ was numerically 
shown to be $1/2$~\cite{Casati95,dima}. 
$\xi$ measured in angular momentum quanta is $\approx t^*$.

\begin{figure}
\begin{center}
\leavevmode
\psfig{figure=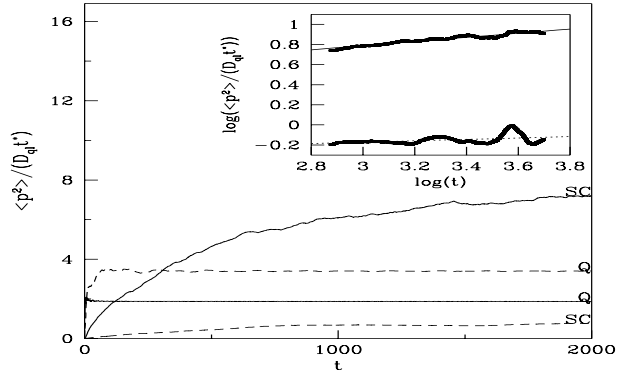,width=3.0in,height=1.9in}
\end{center}
\caption{
Variation of $<p^2>$ (normalized to the saturation value $D_{ql}t^*$)
with the number of kicks $t$. Two representative values of $a$, one
large (Q,$a\approx .61803399$) and the other small (SC,$a \approx 0.042340526$), are
shown for $K=11$ (dashed lines) and $K^*=6.905$ (solid
lines). Note that the larger $a$ value was time-averaged to remove
fluctuations that mask the trends. The inset shows a {\it log-log} plot (at 
longer times, $\approx 5000$ kicks)
for $a=0.042340526$, $K^*=6.905$ (upper curve) and $K=11.0$ (lower curve).
The corresponding slopes are $0.21$ (solid line) and 
$0.06$ (dashed line) respectively.}
\label{fig3}
\end{figure}

Figure.~\ref{fig3} shows $<p^2>/(D_{ql}t^*)$ as a function
of time for two representative values of $a$. For large $a$,
both values of $K$ result in strong saturation with $K=11$ having a 
higher saturation value 
than $K^*=6.905$. However, with decreasing $a$ the enhanced diffusion for
$K^*=6.905$ becomes evident and there is a slowdown in growth in lieu of
saturation (over the time considered). By 
contrast, $K=11$ appears to saturate at a value which 
is about an order of magnitude lower. This difference is
considerably larger than earlier expectations based on quasilinear analysis.
The inset in Fig.~\ref{fig2} displays the existence of power-law behavior in
$<p^2>$ {\it for longer times} at $K^*$ as compared with saturation
at $K=11$. Note that the slope points to sub-diffusive rather than to
super-diffusive growth for which we have no clear explanation at present.

\begin{figure}
\begin{center}
\leavevmode
\psfig{figure=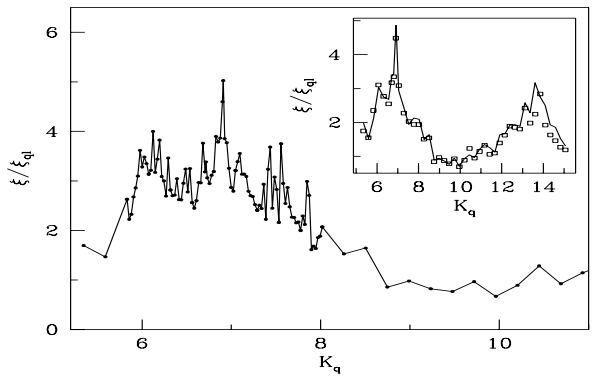,width=3.0in,height=1.9in}
\end{center}
\caption{Oscillations in the localization length $\xi$ normalized to
the quasilinear estimate as a function of $K_q$. $a \approx 0.131267463$ which
makes the corrections to the classical stochasticity parameter 
$K$ significant. The inset shows peaks associated with the
accelerator modes while the main figure
zooms in on the region considered in Fig.1. The inset
also shows $\xi$ computed after $3000$ kicks (points) as well
as $9000$ (line) kicks. Note that the wavefunction averaged over 
$600$ kicks is used to determine $\xi$.}
\label{fig4}
\end{figure}

A stronger signature appears on considering the variation of
localization length $\xi$ with $K$. It was shown~\cite{dima} that
$\xi$ tracks the variations in the classical diffusion coefficient
$D$ when plotted with respect to $K_q=K \sin{(\hbar/2)}/(\hbar/2)$.
It is clear that this can be an important correction for larger values
of $\hbar$. Figure ~\ref{fig4} shows the variation in $\xi$,
normalized to the quasilinear estimate, with respect to $K_q$. $\xi$
is obtained by fitting an exponential to the time-dependent
wavefunction for long times. The peaks in the inset coincide with
the classical accelerator modes at $K=2\pi$ and $4\pi$ and $\xi$
computed at different times reflect the same features. The more
detailed scan displays a reasonable correspondence with the
results shown in Fig.~\ref{fig1}, despite being far from the
semiclassical limit. It should be noted that the highest value
occurs at $K_q \approx K_c$ defined earlier.
 
\begin{figure}
\begin{center}
\leavevmode
\psfig{figure=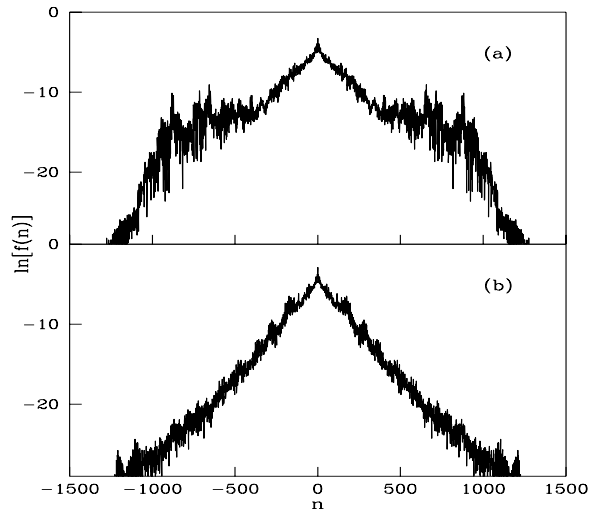,width=3.0in,height=2.5in}
\end{center}
\caption{Lineshapes after $6000$ kicks starting from a plane wave
  initial condition. (a) corresponds to $K_q \approx K_c =
  6.90845\cdots$ while (b) corresponds to $K_q=10.69$. $a \approx 0.131267463$ and
the noise in the lineshapes was reduced by averaging over
$600$ kicks.}
\label{fig5}
\end{figure}

We now consider the nature of the localized wavefunction
when the classical transport is predominantly
anomalous. Figure ~\ref{fig5} shows the evolved
momentum distribution averaged over $600$ kicks. Panel (a)
corresponds to $K_q$ at a peak in Fig.~\ref{fig4} while (b) sits in
the valley. Localization does occur though the characteristic
lineshape is strongly non-exponential for $K_q$ in the neighborhood of
the peaks. The contrast with the clearly exponential  lineshape shown in (b) is
striking. It should be noted that the central region in (a) also displays
curvature and an exponential fit results in a $\xi$ which is considerably 
larger than the quasilinear prediction. However, in case (b), the
computed $\xi$ is well-approximated by the quasilinear
estimate. Moving away from the peak in $K_q$ decreases the prominence of the 
{\it shoulders} in the lineshape though these are still clearly
visible. We also note that the shoulders develop over a time (which
depends on $\hbar$) which is in the neighborhood of the cross-over
time from exponential to power-law behavior shown in Fig.~\ref{fig1}.

\begin{figure}
\begin{center}
\leavevmode
\psfig{figure=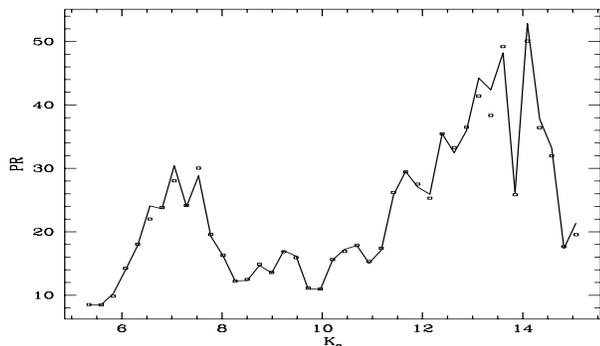,width=3.0in,height=1.8in}
\end{center}
\caption{Participation ratio as a function of $K_q$. Time averages
of the autocorrelation function over $1000$(points) and
$6000$(line) kicks are shown. $a \approx 0.131267463$.}
\label{fig6}
\end{figure}

Arguments for deviations of the localization length from the simple
estimates proceed from the long-time averaged momentum distribution 
$<f_n> = \sum_m |\psi_m(n_0)|^2 |\psi_m(n)|^2$, 
starting from a plane wave $n=n_0$ initial condition. $\psi_m(n)$ is
the plane wave representation of the eigenfunction with quasienergy $\omega_m$.
The analysis of localization assumes that all quasienergy states 
are exponentially localized and that the fluctuations in $\xi$
are small. If this were true in the case of anomalous transport, the
oscillations in Fig.~\ref{fig4} would be directly correlated with the
number of quasienergy states participating in the dynamics. To address
this view, we computed that participation ratio $PR$ from the time
averaged autocorrelation function
\begin{equation}
PR = \lim_{T \rightarrow \infty} \frac{1}{T} |<\psi(T)|\psi(0)>|^2 \;.
\end{equation}
As seen from Fig.~\ref{fig6}, the only correlation with
Fig.~\ref{fig4} occurs in the valley where the lineshape is
exponentially localized. At the peaks, these results strongly
support the idea that anomalous transport results in strong
fluctuations in $\xi$ among the individual quasienergy states.
We have verified this by constructing the `entire' spectrum of 
the quantum map $U$, in the momentum representation, under 
conditions necessary for dynamical localization. The average
value of $\xi$ does not reflect  the large changes in $D$ though
dispersion $\Delta \xi$ exhibits large fluctuations. This view
is being explored further at present. As we argued earlier, a similar
scenario may apply to the interplay of scarring with
dynamical localization.

Theoretically, the classical flights of arbitrary length result (for
the standard map) from a hierarchy of boundary layer 
islands~\cite{george}. The island area $S_n$, associated
period $T_n$ and Lyapunov exponent $\Gamma_n$  scale as: $S_n=\lambda_S^n S_0$,
$T_n=\lambda_T^n T_0$ and $\Gamma_n=\lambda_{\Gamma}^n \Gamma_0$ with the
generation $n$ of the island chain. $\lambda_S <1$, $\lambda_T > 1$
and $\lambda_{\Gamma} \approx 1/\lambda_T$. It is straightforward
to construct a semiclassical break-time appropriate to anomalous
transport using these scaling rules. 

Quantum flights are possible as long as $S_n \geq \hbar$ leading to a critical 
value $n_{\hbar} = | \ln{\hbar/S_0} |/| \ln \lambda_S|$. The corresponding
time is $T_{\hbar}=T_0 (S_0/\hbar)^{1/\mu}$ where
$\mu=|\ln \lambda_S|/\ln{\lambda_T}| >1$ is a classical transport
exponent~\cite{george}. We generalize the  break time $\tau_{\hbar}$ defined earlier
to $\ln{(I/\hbar)}/\Gamma_n$ where $I$ was some characteristic action. 
The longest time corresponds to the smallest
$\Gamma_n=\Gamma_{\hbar}=1/T_{\hbar}$ from which we have
\begin{equation}
\tau_{\hbar}=T_\hbar \ln{(I/\hbar)}= T_0 \left( S_0/\hbar
\right)^{1/\mu} \ln{(I/\hbar)} \;.
\end{equation}
The result points to a power-law dependence on $\hbar$ rather than the
logarithmic one as well as a possible crossover in scaling behavior. Note that the 
saturation time $T_\hbar$ for flights
is a pure quantum effect and one that is necessary for localization,
which cannot occur if arbitrarily long flights exist with high
probability (not exponentially small). Theoretical estimates are
considerably harder when $\hbar > S_0$.

The issue of fluctuations in $\xi$ is relevant to an atom 
optics realization of the kicked pendulum~\cite{expt1}. Here, the
localization properties are similar to those in the quantum
standard map, with the complication that small
islands persist in the classical dynamics. As we have illustrated,
this could lead to strong fluctuations in the localized wavefunction.
More recent experiments have realized the delta-kicked rotor
system~\cite{expt2} where the measurement of diffusion,
over timescales considerably longer than those predicted earlier, can
provide direct evidence of anomalous transport in
select windows of $K$. We note that the signatures of anomalous 
transport are similar to those resulting from external noise, in
both the lineshape and the time evolution of the energy. This may be
especially relevant in the weak noise limit and will be 
considered in detail elsewhere.

In conclusion, we have presented evidence of the role
of anomalous diffusion in enhancing quantum fluctuations.
Note that the standard map is not ideal for isolating the effects of
anomalous transport. The kicked harmonic
oscillator, which results in the classical web map, is
a more interesting candidate where 
super diffusion dominates the classical dynamics~\cite{george}. 
The absence of localization in that system~\cite{Rubaev} and
a recent suggestion for an experimental realization in the
context of ion traps~\cite{zoller} makes this an attractive direction to 
pursue.

The work of BS was supported by the National Science
Foundation and a grant from the City University of New York
PSC-CUNY Research Award Program. GMZ was supported by U.S.
Department of Navy Grants Nos. N00014-93-1-0218 and
N00014-97-1-0426. The authors are grateful to Mark Edelman for 
assistance with the Poincar\'{e} recurrence plot.

\end{document}